\documentclass[11pt]{article}

\usepackage{amsmath,amssymb,amsfonts} 
\usepackage{graphicx}
\usepackage{bm}

\begin{document}

\begin{flushright}
NFPI-11-1
\end{flushright}
\vspace{0.2 cm}

\begin{center}
\Large{\bf On the dynamic nature of charge quantization}

\vspace{0.8 cm}
\renewcommand{\thefootnote}{\fnsymbol{footnote}}
\large{Dennis Crossley}\footnote{dennis.crossley@uwc.edu}

{\emph {Dept.~of Physics, University of Wisconsin-Sheboygan, Sheboygan, WI 53081}}

\vspace{0.8 cm}
\today
\end{center}

\begin{abstract}
It is commonly observed that objects in a gravitational field
experience a rate of acceleration that is independent of their
mass and that, as a result, all massive objects with the same initial conditions follow the same trajectory. It is not generally recognized, however, that charged particles
in an electric field experience an acceleration which is
\emph{inversely proportional} to their mass.  This dynamical
behavior is an interesting clue to the fundamental nature of the electric
force, equally as important as the more familiar behavior of falling bodies, and
seems to be the true significance of the observed fact
that different charged particles have the same magnitude of charge $e$.
\end{abstract}

\noindent
{\emph{Keywords: charge quantization, general relativity, unified field theory 
\\PACS: 04.20.Cv, 12.10.-g, 45.50.-j}} 
\vspace{0.2 cm}

\section{Introduction}
It is curious that different charged fundamental particles have the
same magnitude of electric charge $e$, even though in general they
have different masses.  Electrons and protons, for example, have
masses that differ by nearly a factor of 2000, yet they have the
same magnitude of charge.  This fact is widely recognized, but the
implications of this for the dynamics of charged particles is not. A
comparison between the effects of the gravitational force and the
electric force on the acceleration of different particles sheds some
interesting light on this issue.

\section{Acceleration due to gravity}
Galileo was perhaps the first to observe that all bodies fall at the
same rate.\cite{Galileo} In other words, they all experience the
same rate of acceleration in a gravitational field, independent of
their masses.  This is explained quantitatively by Newton's second
law of motion
\begin{equation}
\label{secondlaw} F=ma,
\end{equation}
and his universal law of gravitation,
\begin{equation}
\label{gravity} F_{grav}=G\frac{M m}{r^2}.
\end{equation}
When the gravitational force in Eq.~(\ref{gravity}) is substituted
into Eq.~(\ref{secondlaw}), the mass $m$ of the falling object
appears on both sides of the equation,
\begin{equation}
\label{samemass} ma=G\frac{M m}{r^2},
\end{equation}
and cancels, giving the result that gravitational acceleration is
independent of the mass $m$ of the falling body,
\begin{equation}
\label{gravacceln} a_{grav}=G\frac{M}{r^2}.
\end{equation}
In the original context of Galileo's observation, namely falling
objects near the Earth's surface, the right side of
Eq.~(\ref{gravacceln}) is simply equal to $g=9.8 \ \mbox{m/s}^2$ and
we get the familiar result that all objects fall with the same
acceleration
\begin{equation}
\label{independent} a_{grav}=g \qquad \mbox{(independent of mass
$m$)}.
\end{equation}
But behind this seemingly simple cancelation of the mass in
Eq.~(\ref{samemass}) lies an unresolved mystery.  The two masses $m$ that
appear in Eq.~(\ref{samemass}) represent conceptually different
quantities.  The mass that appears in equation (\ref{secondlaw}) is
the object's inertial mass, a measure of it's resistance to change
of motion when it is subject to \emph{any} force, not necessarily
gravity. The mass that appears in Eq.~(\ref{gravity}) is the
object's gravitational mass, a measure of its participation in the
gravitational force, whether or not it is being accelerated.
Eq.~(\ref{samemass}) should more correctly read
\begin{equation}
\label{samemass2} m_ia=G\frac{M m_g}{r^2},
\end{equation}
where subscripts $i$ and $g$ have been added to distinguish between
inertial mass and gravitational mass respectively.  Physicists and
philosophers alike have repeatedly pointed out that there is no
fundamental reason why these two masses should be
equal\cite{Jammer1} and the mystery behind this equality is usually
hidden from view by the common practice of using the same symbol $m$
to represent both quantities. That these two masses are proportional
to each other is an empirical observation that was first confirmed
by Newton in his experiments with pendulums\cite{Principia} and
subsequently confirmed with increasing precision by a number of
other physicists.\cite{Jammer2} It is, however, the dynamical
consequence of this equivalence of inertial and gravitational
masses, namely that \emph{the acceleration of a massive object in a
gravitational field is independent of its mass}, which offers a clue
to the nature of the gravitational interaction.

This clue was, in fact, the starting point for Einstein's development of his general theory of
relativity.\cite{Einstein1}  The realization that the motion of bodies
in a gravitational field  is independent of the nature of the bodies
suggested to Einstein that the properties of the gravitational field
could be attributed to the structure of spacetime itself.  The
success of general relativity lies in a mathematical \emph{tour de
force} that tells us how much spacetime must curve in order to
reproduce the observed equivalence of gravitational and inertial
mass.  This is represented mathematically by Einstein's field
equation,
\begin{equation}
\label{genrel} R_{ik} - {\textstyle\frac{1}{2}} g_{ik}R=-kT_{ik}.
\end{equation}
A detailed explanation of this equation, which can be found in any
text on general relativity,\cite{grtexts} need not concern us here.
The only aspect of it that is important for the present discussion
is that the left side describes how much spacetime curves and the
right side describes the distribution of matter which produces this
curvature.  The equations of motion for objects moving in this
curved spacetime then follow from the constraint that they move along
the ``straightest possible path" (a geodesic) through this curved
spacetime.

Einstein's success at ``geometrizing gravity" led to numerous
attempts to extend this approach to other forces.  But while this
geometrization works for the gravitational force, it can not be
extended to other forces, such as the electromagnetic force, for
which the motion of bodies \emph{does} depend on the nature of the
bodies.  Furthermore, general relativity does not explain how (or
why) matter causes space to curve, it only tells us \emph{how much}
space curves.  In this sense, at least, the theory is incomplete.
Einstein, who spent the last three decades of his life searching for
a unified field theory that would describe both gravity and
electromagnetism in terms of a pure field equation, was himself
never fully satisfied with the presence of matter (the right side of
Eq.~(\ref{genrel})) separate from the field (the left side of
Eq.~(\ref{genrel})) in his field equation.  Writing in 1949, over
three decades after he published his theory of general relativity,
Einstein wrote, referring to Eq.~(\ref{genrel}),\cite{Einstein2}
\begin{quote}
The right side is a formal condensation of all things whose
comprehension in the sense of a field theory is still problematic.
Not for a moment, of course, did I doubt that this formulation was
merely a makeshift in order to give the general principle of
relativity a preliminary closed expression.  For it was essentially
not anything more than a theory of the gravitational field, which
was somewhat artificially isolated from a total field of as yet
unknown structure.
\end{quote}

While it is tempting to accept the curvature of spacetime as the
``explanation" for the equivalence of gravitational and inertial
mass, it must be recognized that general relativity does not
answer related questions such as ``\emph{why} do massive objects
cause space to curve in just the right way to preserve this
equivalence of masses?"  and ``why do the other forces behave
differently from gravity?"  Until a theory is advanced which
unifies the gravitational and electromagnetic forces and answers
these questions, it is perhaps best to withhold judgment on
whether general relativity gives us the final word on the
equivalence of masses. Especially when comparing this situation
with that presented by the electric force in the next section, it
may yet be fruitful to continue to ponder the question ``why is
the acceleration of a massive object in a gravitational field
independent of its mass?"

\section{Acceleration due to the electric force}
The situation with the electric force does not suffer from the
confusion between the two masses because the electric force is
proportional to electric charge rather than to gravitational mass,
as is evident in Coulomb's Law,
\begin{equation}
\label{coulomb} F_{elec}=\frac{1}{4 \pi
\epsilon_0}\frac{q_1q_2}{r^2}.
\end{equation}
The details of the source charge(s) are commonly absorbed into the
electric field $E$,
\begin{equation}
\label{coulombE} F=qE.
\end{equation}
This has the advantage that it makes it easier to focus on the
effects of the electric interaction on the charge $q$ without the
distraction of the details of the sources that produce the
electric field. In an electric field $E$, different charged
particles do not accelerate at the same rate as they do in a
gravitational field, but if we consider two different charged
fundamental particles in the same electric field, we observe an
effect which is equally interesting. Let particle 1 have mass
$m_1$ and charge $q_1=e$, and particle 2 have mass $m_2$ and
charge $q_2=e$, where we explicitly use the fact that different
charged fundamental particles have the same unit of electric
charge $e$.  The particles could be, for example, a $\mu^-$ lepton
(charge $-e$) and a $\pi^+$ meson (charge $+e$).  Applying
Newton's second law to both of these particles (ignoring the minus
sign which determines the \emph{direction} of the acceleration),
\begin{equation}
m_1a_1=eE, \qquad m_2a_2=eE,
\end{equation}
it immediately follows that
\begin{equation}
\label{maequalsma} m_1a_1=m_2a_2.
\end{equation}
There is, of course, a long list of fundamental particles with
different masses but the same magnitude of charge
$e$,\cite{particles} and what Eq.~(\ref{maequalsma}) is saying is
that in an electric field they all experience an \emph{acceleration
that is inversely proportional to their mass},
\begin{equation}
a_{electric}\propto \frac{1}{m}
\end{equation}
This dynamic behavior of charged particles in an electric field
seems to be the real significance of the observation that
different fundamental charged particles have the same magnitude of
charge $e$. Nature seems to be giving us a clue here which sheds a different light on the quantization of electric charge.  Instead of asking why different fundamental particles have
the same magnitude of charge $e$, we should be asking
``why is the acceleration of a charged particle in an electric
field inversely proportional to its mass?"

\section{Conclusion}
In the gravitational interaction, it is the fact that the
acceleration of all bodies is independent of their mass that allows
the construction of a theory of gravity in terms of the curvature of
spacetime and independent of the properties of the objects.  In
contrast, the electric interaction causes a different acceleration
for different objects and a similar geometric theory of the electric
force following the pattern of general relativity is not possible.
This, I believe, is the ultimate reason that the attempts of
Einstein and others to develop a unified theory of gravity and
electromagnetism based on the model of general relativity have
failed.

A brief survey of more recent advances in theoretical physics
shows that we are no closer to understanding this difference in
the dynamics of particles in gravitational and electric fields.
The current theory of elementary particles, the standard model,
does not address this issue directly. It assumes the masses and
charges of the elementary particles as inputs to the theory but
cannot account for the observed mass spectrum and
sheds no light on why different charged particles have the
same magnitude of charge. String theory offers an enticing suggestion that
different masses correspond to different resonances of a
fundamental string, but the mathematical formalism is not yet
developed enough to allow the calculation of the mass spectrum
from fundamental principles, and like the standard model also
offers no clue as to why different elementary particles have the
same charge. And finally, active research into quantum gravity
seems directed more toward understanding the properties of
spacetime itself than toward the behavior of the objects that
populate it.

Nature is offering us clues to a deeper
understanding of the fundamental particles and their interactions, and
we would do well to pay attention to them.  The two facts that (1)
gravitational acceleration is independent of mass and (2) electric
acceleration is inversely proportional to mass are clues that should
be given equal importance.  Add to these the equally unexplained
mystery of why gravity is always attractive but the electric force
can be both attractive and repulsive, and we get a sense of the
limits of our understanding of these fundamental forces.  Any theory
which attempts to unify gravity and electromagnetism must account
for these differences in behavior.  Existing theory has not been able to resolve this issue.  We need a new perspective, a new way of thinking about this issue.  Instead of thinking of charge quantization as a particle property, we should be thinking of it as a consequence of the dynamics of particle interactions.
Instead of asking ``why do different charged particles
have the same magnitude of charge $e$?" we should be asking ``why
do different charged particles experience accelerations that are
inversely proportional to their masses?"  


\end{document}